# Negotiation in collaborative assessment of design solutions: an empirical study on a Concurrent Engineering process

Géraldine Martin
AEROSPATIALE MATRA AIRBUS
BTE/SM/CAO M0101/9 316 route de Bayonne 31060 Toulouse cedex 03, France
Françoise Détienne
Eiffel Group "Cognition and Cooperation in Design" INRIA
Domaine de Voluceau, Rocquencourt, BP 105, 78153 Le Chesnay, France
Elisabeth Lavigne
AEROSPATIALE MATRA AIRBUS
BTE/SM/CAO M0101/9 316 route de bayonne 31060 Toulouse cedex 03, France

**Abstract**

In Concurrent engineering, design solutions are not only produced by individuals specialized in a given field. Due to the team nature of the design activity, solutions are negotiated. Our objective is to analyse the argumentation processes leading to these negotiated solutions. These processes take place in the meetings which group together specialists with a co-design aim.

We conducted cognitive ergonomics research work during the definition phase of an aeronautical design project in which the participants work in Concurrent Engineering. We recorded, retranscribed and analysed 7 multi-speciality meetings. These meetings were organised, as needed, to assess the integration of the solutions of each speciality into a global solution.

We found that there are three main design proposal assessment modes which can be combined in these meetings: (a) analytical assessment mode, (b) comparative assessment mode (c) analogical assessment mode. Within these assessment modes, different types of arguments are used. Furthermore we found a typical temporal negotiation process.

## 1 Introduction

In the face of growing competition, firms are increasingly having to reconsider their organisation and their processes. The development of methodologies supporting collective (or team) work is one key to success: Concurrent Engineering (CE), assumed to be one solution to achieve greater efficiency in the collective design process. The CE model prescribes various phases of design together with their temporal organisation. It consists in a "systematic, integrated and simultaneous" way of developing products and associated processes, in particular manufacturing processes, e.g., production [2]. Based on this model, methods and tools aim at guiding the organisation of the design, both at individual and team level.

Some companies have implemented CE in the course of re-engineering of their design processes. To do so, they extensively deploy tools for design support and for technical data management. In one aeronautical company, this change is being supported by research in Cognitive ergonomics. It is in this framework that a field study has been performed.

In Concurrent engineering, design solutions are not only produced by individuals specialized in a given field. Due to the team nature of the design activity, solutions are negotiated [1]. Our objective is to analyse the argumentation processes leading to these negotiated solutions. These processes take place in the meetings which group together specialists with a co-design aim.

We conducted cognitive ergonomics research work during the definition phase of an aeronautical design project in which the participants work in Concurrent Engineering. 10 different specialities are involved. We recorded, retranscribed and analysed 7 multi-speciality meetings. These meetings were organised, as needed, to assess the integration of the solutions of each speciality into a global solution.

After a brief presentation of our theoretical framework and hypotheses, we present an empirical study aimed at analysing the negotiation in an industrial Concurrent Engineering context. Our approach is strongly oriented by cognitive ergonomics work on the notion of constraint, and linguistics work on argumentation.

## 2 Theoretical framework and hypotheses

For design problems, the solutions are not unique and correct but various, and more or less satisfactory according to the constraints that are considered. The designers assess the solutions they develop according to their own specific constraints, which reflect their own specific points of view, in relation with the specificity of the tasks they perform and their personal preferences ([6],[7]). Also assessment modes may vary and involve more or less explicit constraints.

Constraints are cognitive invariants which intervene during the design process. The notion of constraints has been understood from different angles (1) according to their origin - prescribed constraints, constructed constraints, deduced constraints, (2) according to their level of abstraction, and (3) according to their importance – validity constraints and preference constraints ([3][6][8]).

Up to now, the assessment of design solutions has been mostly studied in the individual design process. In design activities, the assessment intervenes (1) to appreciate the suitability of partial solutions to the usual state of resolution of the problem, and (2) to select one of the solutions envisaged ([2][3][4]). The finality of this assessment is to make the decision to change one of its components, or to pursue the design if the assessment is positive[5].

Previous studies on individual design [4] have shown that various kinds of assessment modes may be involved. Bonnardel distinguishes between the following three assessment modes:
(a) **analytical assessment mode**, i.e., systematic assessment according to constraints,
(b) **comparative assessment mode**, i.e., systematic comparison between alternative proposed solutions and,
(c) **analogical assessment mode**, i.e., transfer of knowledge acquired on a previous solution (accepted or not) in order to assess the current solution.

In collective design, we make the assumption that similar assessment modes may be found. With respect to linguistic work on argumentation ([9][10]), we will consider that these modes involve, in a meeting situation, the use of various types of arguments.

In the collective assessment, different specialities are going to be present, and they are going to have to justify their design choice so they are going to produce arguments. The purpose of these arguments is to provide information to convince the other people of the pertinence and veracity of the information provided in order to tend towards a conclusion that pushes them towards accepting the proposal [9]. When everyone has a joint will to reach agreement, we shall talk about negotiation. Negotiation does not force a person to accept a solution, dialogue makes it possible to go towards one conclusion rather than another, i.e., for example, the conclusion can be a compromise between what each person wants.

Linguists distinguish different kinds of arguments , argument by comparison, argument by analogy, argument of authority.

**argument by comparison**
argument by comparison compares several objects in order to assess them in relation to each other. Comparisons can be made by opposition, by classification and quantitative classification.

**argument by analogy**
These are arguments that highlight a precedent, i.e., they enable the present case to be compared to a typical case proposed as a model.

We consider that the comparative assessment mode and the analogical assessment mode may involve what linguists call argument founded on an example, argument by comparison or argument by analogy. Most of these arguments can take the status of argument of authority depending on factors which give a particularly strong weight to the argument.

**argument of authority** is an indisputable argument which is built on a quotation of statements, so it is in no way proof, even if it is presented as such. In general, the proposer's argument is the fact that it has been expressed by a particular authorized person, on whom he relies, or behind whom he hides.

Furthermore, due to the collective nature of the assessment process, we expect to observe combined assessment modes: in this case, each participant in an assessment meeting may use one or several assessment modes in order to convince the other participants.

Our research questions are :
(1) whether such combined assessment modes occur in assessment meetings;
(2) if so, whether there is a typical temporal organization of these assessment modes (or temporal negotiation patterns).

## 3 Methodology

### 3.1 Context

We conducted a field study on a design project in an aeronautical company. The project actors followed a CE methodology. The goal of the project was to design a new aircraft. The total duration of this project was three years. The study focused on the definition phase for the design of the aircraft centre section. This phase involved nine various fields of expertise for a total of approximately 400 actors. These actors use computer aided design (CAD) and product data management (PDM) to support design.

All the specialities work on the same part of the aircraft but each person according to his technical competence. "Informal" inter-speciality meetings are organized, as needed, to assess the integration of the solutions of each speciality into a global solution.

This research work involved seven meetings representing a representative sample of the meetings observed in the integrated design group. These meetings last between 15 minutes and an hour. We recorded two types of meetings :
- **meetings between Design Office (D.O) specialities.** We have a configuration of meetings between designers in different fields, in which a D.O field (structure speciality) presents the same problem to five other D.O. fields (system installation speciality). We thus have an invariant in the structure solution proposed.

- **meetings between D.O. specialities** (structure speciality) **and specialities which traditionally intervene later in design** (production and/or maintenance).

### 3.2 Collection of data

We took part in 7 of these meetings as observers. On the basis of audio recordings and notes taken during the meeting, we retranscribed the full content of the meetings. We also conducted interviews afterwards with the various participants to validate the coding we had made of them and make explicit a certain amount of information that was implicit in the meetings.

### 3.3 Coding scheme

The protocols resulting from the retranscriptions were broken down according to the change of locuters. Each individual participant utterances corresponds to a "turn". Each turn was coded according to the following coding scheme and broken down again as required to code finer units.

Our coding scheme comprises two levels :
- **a functional level** : it highlights the way in which collective design is performed. Each unit is coded by a mode (request/assertion) an action (e.g., assess) and an object (e.g., solution n). At this level, a turn can be broken down into finer units according to whether there is a change in mode, activity or object.
- **an argumentative level** : the aim here is to bring out the structure of the speech on the basis of a dialogue situation.

We coded the proposals for solutions made and the different types of arguments used by the speakers during the meetings.

## 4 Results

We found that there are three main design proposal assessment modes which can be combined in these meetings: (a) analytical assessment mode, (b) comparative assessment mode (c) analogical assessment mode. Within these assessment modes, arguments presented to defend a proposal for solution may take the status of " argument of authority ". Furthermore, we found a typical temporal negotiation pattern.

### 4.1 A general model of the assessment process

For the seven meetings analysed, whatever the problem involved, a solution is proposed by a speciality M1. This solution is called the initial solution. M1 will give arguments to support it in order to convince the other speciality, M2 (or the other specialities when more than two specialities are present). This solution may be accepted immediately by M2 who is convinced of the pertinence of the solution. On the other hand, M2 could refuse it, which is the most frequent case. Then follows a negotiation between the two specialities in order to reach a consensus. However, sometimes the negotiation fails and M1 and M2 must then find a compromise. An alternative solution is then proposed by M1 or M2 which will in turn be assessed. Often, several alternative solutions are proposed before a negotiated solution is reached. Finally, it sometimes happens that the meeting does not enable a result to be achieved. Each speciality must then work again before another meeting is convened

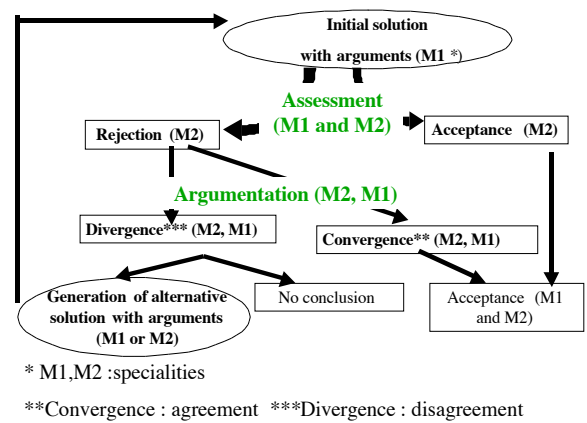

\* M1,M2 :specialities

\*\*Convergence : agreement  \*\*\*Divergence : disagreement

**Figure 1 : The assessment process**

### 4.2 Combined assessment modes

The first type of result involves the way in which the proposals for solutions are assessed during these meetings. We have revealed the existence of analytical, comparative or analogical assessment modes in these meetings. This type of result is similar to the assessment modes analysed in individual design [4].

In addition, we have highlighted combined assessment modes, e.g. analytical/analogical. We present these modes and illustrated them graphically through examples.

#### 4.2.1 *analogical /analytical assessment*

This mode combines analogical assessment and analytical assessment. In the framework of analogical reasoning, the current solution (the one which is proposed for evaluation) is called the target solution whereas the analogical solution (a previous solution which is brought up in the argumentation process) is called the source solution.

Figure 2 illustrates graphically such a combined assessment mode. In this example, specialists M1 use the analogical/analytical assessment to convince specialists M2 to accept the solution S1 proposed by M1.

Specialists M1 propose a solution, the target solution S1, which is rejected by specialists M2. In order to convince specialists M2 of the adequateness of S1, specialists M1 make reference to an analogical solution, the source solution S2. S2 is a solution which was accepted in a past context. In this context S2 was a solution negotiated between M1 and M3 : even if this solution was not so easy to use by specialists M3 (this solution was not ideal in terms of some constraints important for these specialists), they finally accepted it. In their argumentation, specialists M1 analyse the source solution S2 according to a set of constraints (analytical assessment). They make explicit positive arguments as well as negative arguments and defend the idea that the specialists M3 were able, in the past, to accept this evaluation and therefore the source solution S2. The conclusion of this negotiation process is the acceptance, by M2, of the target solution S1.

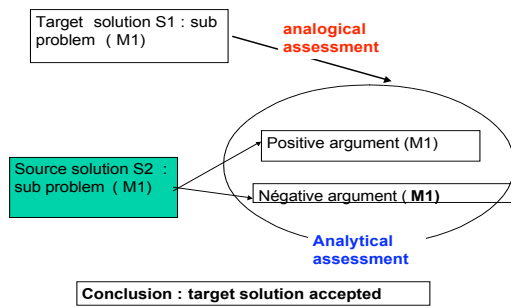

**Figure 2 : Analogical /analytical assessment**

### 4.2.2 *comparative/analytical assessment*

This mode combines comparative assessment and analytical assessment. The comparative assessment mode involves systematic comparison between the current solutions and one or several alternative proposed solutions. These solutions are alternative to the current proposed solution (the one originally to be assessed).

Figure 3 illustrates graphically such a combined assessment mode. In this example, each specialist will propose his own alternative solution. None of them accept the current proposed solution.

Specialists M1 propose an alternative solution Salt 1 whereas specialists M2 propose another alternative solution : Salt 2. Each alternative solution is then analytically analyzed by participants of both specialities. Specialists M1 positively assess Salt1 (their own proposed alternative solution) and negatively assess Salt2. Conversely, specialists M2 positively assess Salt2 (their own proposed alternative solution) and negatively assess Salt1. These analytical assessments allow each specialist to compare the suitability of the two alternative solutions according to various design constraints. In doing so, each speciality makes explicit the design constraints which are judged more important in his field. The conclusion of this negotiation process is that neither of the two proposed alternative solutions are accepted. Rather, a third alternative solution, which is a compromise between Salt1 and Salt2, is generated.

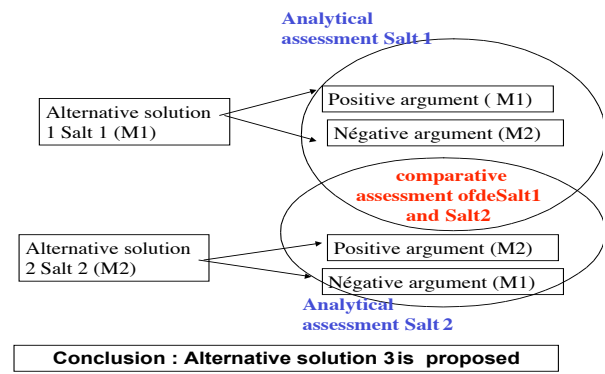

**Figure 3 : Comparative/analytical assessment**

### 4.2.3 *Comparative/analogical assessment*

This mode combines comparative assessment and analogical assessment. Figure 4 illustrates graphically such a combined assessment mode. In this example, specialists will propose an alternative solution (comparative assessment) and will defend this solution in reference to a previous source solution which was accepted in the past (analogical assessment).

Specialists M1 propose and defend the current solution S1. Specialists M2 propose an alternative solution Salt1. In order to defend this alternative solution, they make reference to a source solution, accepted in a past context, which is analogical to Salt1.

This source solution is then analogically assessed by the different specialists. This evaluation allows the specialists to compare the advantages (positive arguments) and drawbacks (negative arguments) of the current solution S1 and its alternative solution Salt1.

Specialists M1 give negative arguments toward Salt1 based on negative arguments toward the source solution ; this allows them to show, by comparison, the advantages of solution S1. Conversely, specialists M2 give positive arguments toward Salt1 based on positive arguments toward the source solution ; this allows them to show, by comparison, the drawbacks of solution S1. The conclusion of this negotiation process is the absence of any negotiated solution or any consensus. In fact, due to the disagreement between the specialists on the source, a task is planned in order to verify information related to the source solution. The design rationale about the source solution has to be reconstructed for the next meeting.

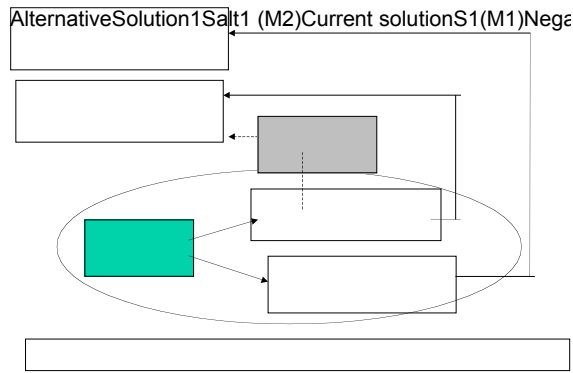

**Figure 4 : Comparative/analogical assessment**

### 4.3 Argumentation

Argumentation means provoking or increasing other people's support of the theories submitted to then for agreement. For argumentation to be effective, the designers, in the collective assessment use different type of arguments according to an order which seems to have become customary. Arguments used are of different nature. Due to the nature of the task, a design task, many arguments explicitly or implicitly reference make of design constraints. Furthermore, arguments can take the status of argument of authority depending on specific factors.

#### 4.3.1 <u>Use of Contraints</u>

Arguments enabling a design solution proposal to be defended are often characterized by the use of constraints.

Constraints can be explicit or implicit in the argument as it is expressed by a speaker. The implicit or explicit nature can depend on the postulate of shared knowledge made by the speaker. Now this postulate is not always confirmed.

An argument often covers not only the explicit constraint but a hierarchical network of implicit constraints; this network can be broken down in order to convince the other participants.

We have observed that specialists firstly make explicit the constraints at the top of the hierarchy and, when it is necessary to bring other arguments to convince the other specialists, then the constraints lower in the hierarchy are made explicit.

We observed that the same constraint (the same terms are used by different speakers) can have different meanings according to the speakers, and more specifically, according to the speciality of the speaker.

In this case it is necessary to distinguish the two slopes of the sign, the signifier and the meaning. The meaning can have the same generic seme for different speakers but very different functional seme. For example, a cost constraint can, for one speciality, mean "production cost" and, for another speciality, mean "design cost". It seems particularly true for general constraints prescribed for all the actors of the design process (e.g., the cost) as opposed to constraints derived by a speciality (e.g., structure).

We also found that constraints can be weighted differently according to specialities. There is no absolute weighting except for certain constraints (1). The weighting is performed in a context according to the type of problem considered. One assumption is that the weighting is done not on the constraint as such but on its meanings. The context of the problem would make it possible to select a particular meaning.

#### 4.3.2 <u>Use of argument of authority</u>
Any argument can take the status of argument of authority depending on specific factors of the situation. This argument is presented as inconstestable and therefore it has a particularly strong weight in the negotiation process.

We have found that an argument can take the status of argument of authority depending on :
- the status, recognised in the organisation, of the speciality that expresses it.

- the expertise of the proposer. The argument is going to make reference to a person recognized by all to be an expert in the speciality. It will be something like *" It's Alphonse who said it would be more logical like that to pick up on these parts of the stringers"*.

- the "shared" nature of the knowledge to which it refers. This is typically the case in analogical assessment, when the participants in a meeting have shared knowledge about the source solution, e.g., everybody agrees that it works in this similar context. In some cases, we observed that participants do not share knowledge about the source (as in 4.2.3).

### 4.4 Temporal negotiation patterns

As explained before, we have found that combined assessment modes occur in assessment meetings.

---

[1] Each speciality has some specific strong constraints : for the structure specialists, for example, there are weight constraint and structure constraint.

Another research question was whether there is a typical temporal organization of these assessment modes. We found that different assessment modes are used in the order shown in Figure 5:
- Step1: Analytical assessment mode of the current solution;
- Step 2: if step 1 has not led to a consensus, comparative or/and analogical assessment is involved;
- Step 3: if step 2 has not led to a consensus, one (or several) argument(s) of authority is(are) used.

Firstly the current solution is assessed. This is made using an analytical assessment mode. Arguments used by the two (or more) specialities may use more or less explicit design constraints. Specialists M1 use arguments to convince M2 and M2 does the same thing. Based on this analytical assessment, a consensus can be found and negotiation is finished.

If no consensus has been found, then either M1 or M2 (or more rarely both) use either an analogical assessment mode or a comparative assessment mode of the solution. The two types of assessment may also be combined. This can lead again to a consensus toward the initial solution or toward a proposed alternative solution.

If no consensus has been found, either M1 or M2 propose one or several arguments of authority. This generally leads to a consensus.

An example of a non converging negotiation process was illustrated in 4.2.3 (Figure 4). In this particular case, each specialist had different arguments related to the same source. The use of the source could have led to a consensus based on the shared knowledge concerning the source (argument of authority). In this particular case, this process was disrupted and no consensus could be found.

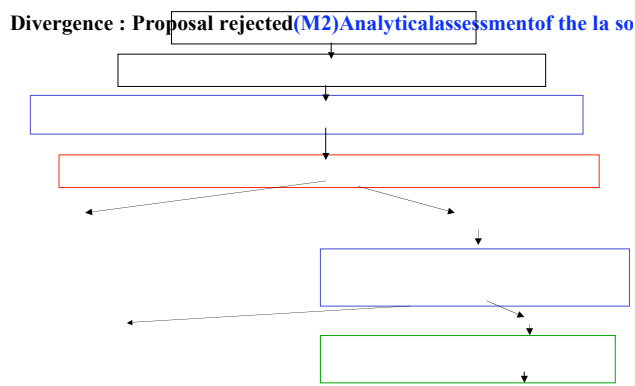

**Figure 5 : The Argumentation process**

## 5    Conclusion

To sum up, we have found that there are three main assessment modes which can be combined in design assessment meetings. Within these assessment modes, different types of arguments are used. Furthermore we found a typical temporal negotiation process.

Two courses of action are now being studied. The first is to improve design rationale traceability. Indeed, only a part of this design rationale is now absent from the minutes of meetings. The second is capitalization of the knowledge brought into play in the logical assessment and analogical /analytical assessment. This knowledge is associated with particular problems encountered in the past and procedural type general knowledge. This capitalization would be done for reutilization purposes.

Our long-term objective is to analyse and support the integration of points of view in multi-speciality design in order to improve the search for a compromise between designers in design reviews. Indeed, it is in assessment meetings that we can observe the confrontation of the points of view of the various participants in design. Owing to the collective nature of the activity, points of view are expressed, more or less explicitly, through argumentation (Plantin,1996).